\documentclass[twocolumn,10pt,trackchanges]{aastex63}

\received{January 7, 2021}
\accepted{June 2, 2021}
\submitjournal{AJ}

\shorttitle{}
\shortauthors{Ito et al.}
\graphicspath{{./}{figures/}}

\usepackage{amsmath}
\usepackage{comment}
\usepackage{threeparttable}
\usepackage{bm}

\begin{document}

\title{Spatial variations of magnetic field along active galactic nuclei jets on sub-pc to Mpc scales}

\correspondingauthor{Soichiro Ito}
\email{s-ito2@akane.waseda.jp}

\author{S. Ito}
\affiliation{Faculty of Science and Engineering, Waseda University, 3-4-1, Okubo, Shinjuku, Tokyo, 169-8555, Japan}

\author{Y. Inoue}
\affiliation{Department of Earth and Space Science, Graduate School of Science, Osaka University, Toyonaka, Osaka 560-0043, Japan}
\affiliation{Interdisciplinary Theoretical and Mathematical Science Program (iTHEMS), RIKEN, Saitama 351-0198, Japan}
\affiliation{Kavli Institute for the Physics and Mathematics of the Universe (WPI), The University of Tokyo, Kashiwa, Chiba 277-8583, Japan}

\author{J. Kataoka}
\affiliation{Faculty of Science and Engineering, Waseda University, 3-4-1, Okubo, Shinjuku, Tokyo, 169-8555, Japan}
 

\begin{abstract}
  We report the systematic analysis of knots, hotspots, and lobes in 57 active galactic nuclei (AGNs) to investigate the variation of the magnetic field along with the jet from the sub-pc base to the terminus in kpc-to-Mpc scales. 
  Expanding the number of radio/X-ray samples in \citet{2005ApJ...622..797K}, we analyzed the data in 12 FR~I and
  30 FR~II radio galaxies, 12 quasars, and 3 BL Lacs that contained 76 knots, 42 hotspots, and 29 radio lobes.
  We first derived the equipartition magnetic fields in the cores and then estimated those in various jet components by assuming $B_{\rm est}$ $\propto$ $d^{-1}$, where $d$ is the distance from the jet base. 
  On the other hand, the magnetic field in large-scale jets (knots, hotspots, and lobes), $B_{\rm eq}$, can be estimated from the observed flux and spatial extent under the equipartition hypothesis. 
  We show that the magnetic field decreases as the distance along the jet increases, but generally gentler than $\propto d^{-1}$. 
  The increase in $B_{\rm eq}/B_{\rm est}$ at a larger $d$ may suggest the deceleration of the jet around the downstream, but there is no difference between FR~I and FR~II jets. 
  Moreover, the magnetic fields in the hotspots are systematically larger than those of knots and lobes.  
  Finally, we applied the same analysis to knots and lobes in Centaurus~A to check whether
  the above discussion will hold even in a {\it single} jet source.

\end{abstract}



\section{Introduction} \label{sec:intro}
Over a century has passed since the discovery of the M87 jet in 1918 \citep{1918PLicO..13....9C}. However, the nature of jets in active galactic nuclei (AGNs) is still one of the biggest mysteries in astrophysics. 
After the launch of the jets from the vicinity of central supermassive black holes (SMBHs), AGN jets are believed to transport a huge amount of energy into the interstellar and intergalactic space, either in the form of kinetic and Poynting fluxes. Such interaction may suppress the star-formation activity in galaxies, the so-called ‘‘AGN feedback’’  \citep{1998A&A...331L...1S,2006MNRAS.365...11C, 2012ApJ...745L..34Z,2012ARA&A..50..455F, 2014A&A...562A..21C,2020A&ARv..28....2V,2021NatAs...5...13L}.
Therefore, unveiling the physical mechanism of the AGN jet is essential for understanding galaxy formation.

The issues of AGN jets yet to be resolved are categorized into formation, collimation, particle composition, and acceleration.
Here, the magnetic fields along the jets are critical parameters for understanding these issues \citep{2012SSRv..169...27P, 2015SSRv..191..441H}. For example, at the jet base, the theoretically plausible explanation for the jet launch requires the support of the magnetic fields threading the central SMBH to extract the jet power from rotating SMBHs \citep{1977MNRAS.179..433B}. Although various numerical simulations confirm this process as a plausible and efficient jet power extraction mechanism \citep{1985PASJ...37..515U, Komissarov2007MNRAS.380...51K, Tchekhovskoy2010ApJ...711...50T, Tchekhovskoy2011MNRAS.418L..79T, McKinney2012MNRAS.423.3083M, Takahashi2016ApJ...826...23T}, this process has not been observationally validated owing to the limited spatial resolution of observations. In the sub-pc scales, near the jet core, blazar observations indicate the dominance of particle energy in the jet, even though the Poynting flux dominates the jet at the base \citep[e.g.,][ but see also, e.g., \citet{Dermer2015ApJ...809..174D, Tavecchio2020MNRAS.491.2198T}]{Tavecchio1998ApJ...509..608T, Tavhecchio2010MNRAS.401.1570T, Inoue2016ApJ...828...13I}. In addition, according to the compilation of the observations of kpc jets of Centaurus~A  \citep[Cen~A;][]{HESS2020Natur.582..356H}, magnetic fields in the knots at kpc scales are likely to be amplified, while the entire diffuse jet is weakly magnetized \citep{2020ApJ...901L..27S}. 
Furthermore, in the hotspots and lobes at the terminal of the jet, the particle contents, such as protons, have also been investigated from the magnetic field by applying the inverse Compton model \citep{2005ApJ...626..733C,2004ApJ...612..729H,2018MNRAS.476.1614C}.

The spatial distribution of the magnetic fields along the jet axis is an essential factor for revealing the dynamics of jets. While systematic differences in the magnetic field in the jet knots, hotspots, and radio lobes (hereinafter collectively referred to as ‘‘jet components’’) has been investigated in literature \citep[e.g.,][]{2005ApJ...622..797K, 2005ApJ...626..733C}, only a few attempts have been made to systematically link the physical quantities of the jets from upstream to downstream.
Recently, \citet{2019ApJ...878..139T} confirmed that the magnetic field tends to decay inversely proportional to distance from the core by comparing two knots in Cen~A with other $\gamma$-ray detected radio galaxies.
This indicates how the magnetic field strength of the relativistic jet scales with the distance from the central black hole, which may shed new light on the mystery of the AGN jet.

This study aims to establish the relationship between the magnetic field and the distance from the central SMBHs using 57 AGNs. Moreover, we investigate a nearby radio galaxy Cen~A in detail, which has a bright core in the sub-pc, and the jet components are resolved from kpc jet knots to sub-Mpc lobes.  In $\S$~2, we describe how we estimate the volume and distance from the jet base, and summarize the physical properties of the cores and jet components.
In $\S$~3, we present the two methods for magnetic field estimations. 
In $\S$~4, we discuss the relationship between the magnetic field and the distance along with the jet. Our conclusion is summarized in $\S$~5.
Throughout this paper, we have used standard cosmology terms, with $H_{\rm 0}=71\mathrm{km\  s^{-1}Mpc^{-1}}, \Omega_{\rm m}=0.27$, $\Omega_{\rm \Lambda}=0.73$.
\newpage


\section{DATA}
\label{sec:DATA}
The XJET catalog \citep{2002ApJ...565..244H}\footnote{\url{https://hea-www.harvard.edu/XJET/}} provides information on 117 spatially extended AGN jets including X-ray and radio references. Our aim was to investigate the spatial variation of the magnetic fields along with AGN jets. Therefore, from the XJET catalog, we selected AGNs 1) that have at least one or more knots, hotspots, and lobes detected both in radio and X-ray bands, and 2) that have VLBI observations toward the cores and downstream jet components. The first restriction is to study the radial dependence, whereas the second restriction is to ensure the availability of radio fluxes of both the core and jet components. The XJET catalog also contained a blazar population. For the blazar population, we further restrict samples 3) that have apparent velocity information,  because they are strongly beamed objects. After applying these selection criteria, our sample contains 57 AGNs, i.e., 12 FR~I radio galaxies (FR~Is), 30 FR~II radio galaxies (FR~IIs), 12 quasars (QSOs), and 3 BL Lac objects (BL Lacs). QSOs contain nine flat-spectrum radio quasars and three steep-spectrum radio quasars. Table \ref{tab:core_data} summarizes the object name, classification, redshift $z$, and reference information. Note that the XJET catalog is not a complete catalog of the sky because it summarizes the various pointing observations available in the literature. Therefore, our sample may be incomplete. The e-ROSITA survey results can provide a complete X-ray jet catalog.

\begin{figure}
    \centering
    \includegraphics[keepaspectratio, width=\linewidth]{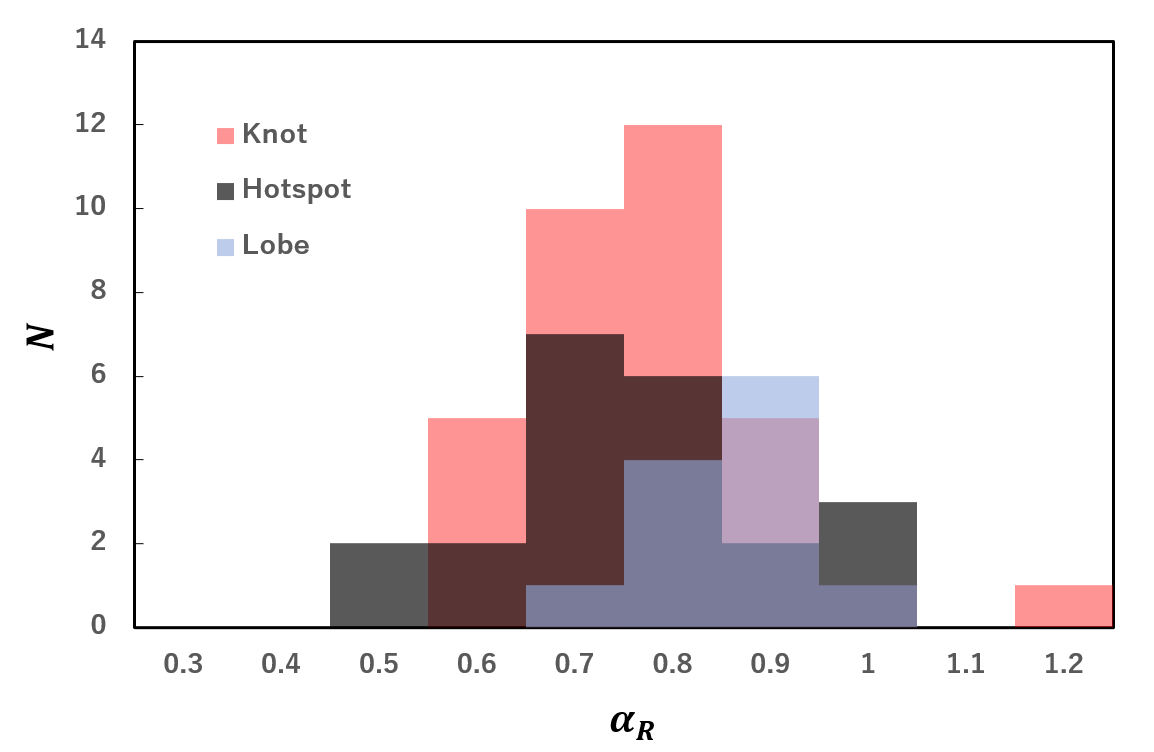}
    \caption{Distribution of the radio spectral index $\alpha_{\rm R}$ for knots, hotspots, and lobes.}
    \label{fig:radio_index}
\end{figure}

Table \ref{tab:comp_data} lists the observational properties of the cores and jet components of the selected AGNs, which contain 76 jet knots, 42 hotspots, and 29 radio lobes.
In this paper, we classified the jet components following the definition in the referenced literature. Briefly, ``knots'’ are defined as distinct structures in the jet, ``hotspots'' are compact structures located at the terminal of the jet, and ``lobes'' are diffuse structures located at the terminal of the jet.
As shown in Figure \ref{fig:radio_index}, the radio spectral index $\alpha_{\rm R}$ of knots, hotspots, and lobes has relatively a narrow distribution with peaks at $\sim$ 0.75. When we cannot obtain $\alpha_{\rm R}$, we assume $\alpha_{\rm R}=0.75$ for both the core and jet components \citep{2005ApJ...622..797K}.

Table.~\ref{tab:comp_data} also summarizes radio flux densities at 5~GHz, i.e., $f_{\rm 5}$, of all components, which we collected from the literature, as listed in Table~\ref{tab:core_data}. For the cores, we also refer to the 3CRR database\footnote{\url{https://3crr.extragalactic.info/cgi/database}}, which contains the latest VLA observations for AGNs summarized by \citet{1983MNRAS.204..151L}. When $f_5$ is unavailable from the literature, we estimate it using the available radio flux at the nearest frequency as well as $\alpha_{\rm R}$. This extrapolation will not induce significant uncertainties in our analysis because the magnetic fields depend on $f_5^{2/7}$ (See, Eq \ref{eq:B_eq2}).

We set the viewing angle of $\theta_{\rm inc}=30^\circ$ for two FR~Is, 15 FR~IIs, and one QSO because their $\theta_{\rm inc}$ are unavailable.
Table \ref{tab:comp_data} lists the Doppler beaming factor $\delta$.
For the core, we derived $\delta$ directly from $\theta_{\rm inc}$ and the apparent velocity $\beta_{\rm app}$ at the inner jet, that is, $\delta=\Gamma^{-1}(1-\beta \cos \theta_{\rm inc})^{-1}$, where $\beta=\beta_{\rm app}(\sin \theta_{\rm inc}+\beta_{\rm app}\cos \theta_{\rm inc})^{-1}$ and $\Gamma =1/\sqrt{1-\beta^2}$.
Because $\beta_{\rm app}$ is unavailable for some FR~I/FR~IIs, $\delta$ of these AGN cores were set to $\delta = 1$.
We then assume the same $\delta$ as applied to the other downstream knots. For the hotspots, both the observations and simulation studies suggest that the plasma velocity can be mildly relativistic, i.e., a few $\Gamma$ at most \citep{1999ApJ...523L.125A,2003ApJ...589L...5G,2007ApJ...662..213S}. Therefore, to provide a conservative estimate, we set $\delta = 3$ as an ultimate case; however, the same results were obtained if we set $\delta = 1$. Finally, we set $\delta = 1$ for the radio lobes, as suggested by \citet{2005ApJ...622..797K}.

We also need the geometrical information of each component, such as the angular size and distance. Because the cores were not resolved for most of the sources, we assumed a spherical geometry with a radius of $R=10^{17}$cm. This radius corresponds roughly to $R\sim 3\times 10^2 r_{\rm s}$, where $r_{\rm s}$ is the Schwarzschild radius of AGN with a black hole mass of $10^9 M_{\rm \odot}$. The corresponding angular sizes $\theta$ of the cores are listed in Table \ref{tab:comp_data}. We set the cores located at $d_{\rm core} = R/\sin \theta_{\rm op} \sim R\Gamma$ from the central black hole, where $\theta_{\rm op}$ is the opening angle of the jet. When $\theta_{\rm op}$ is available \citep[Pictor~A, Cen~A, and M87;][]{1999Natur.401..891J,2006PASJ...58..211H,2016MNRAS.455.3526H}, $d_{\rm core} = R/\sin \theta_{\rm op}$ is adopted. For the other sources, we use $d_{\rm core}=R\Gamma$. 

As the sizes of the jet components, we assume a spherical geometry \citep{2005ApJ...622..797K} for those that are not spatially resolved in both radio and X-rays (57 $\%$ of the total jet components). For these unresolved components, we set the minimum spatial resolution of the telescopes as the angular size. For example, the Chandra X-ray satellite achieved $0.3^{\prime\prime}$. For the resolved sources, X-ray images confirm the validation of the spherical approximation, and we applied the measured angular sizes. However, 11 radio lobes showed nonspherical geometries. We approximate their geometry as a cylinder or rotational ellipse following \citet{2005ApJ...626..733C} and derive their volumes. We then estimated their sizes in terms of spherical geometry using the estimated volumes. All size information is summarized in Table~\ref{tab:comp_data}.

The distance from the core, $d$, for each jet component is
calculated as $d=d_{\rm L}\theta_{\rm D}(1+z)^{-2}/\sin\theta_{\rm inc}$, where $d_{\rm L}$ is the luminosity distance and $\theta_{\rm D}$ is the observed angular distance from the jet base to the jet components.


\section{Modeling and results} \label{sec:Modeling}

\begin{figure*}[t]
    \begin{center}
        \includegraphics[keepaspectratio,scale=1.3]{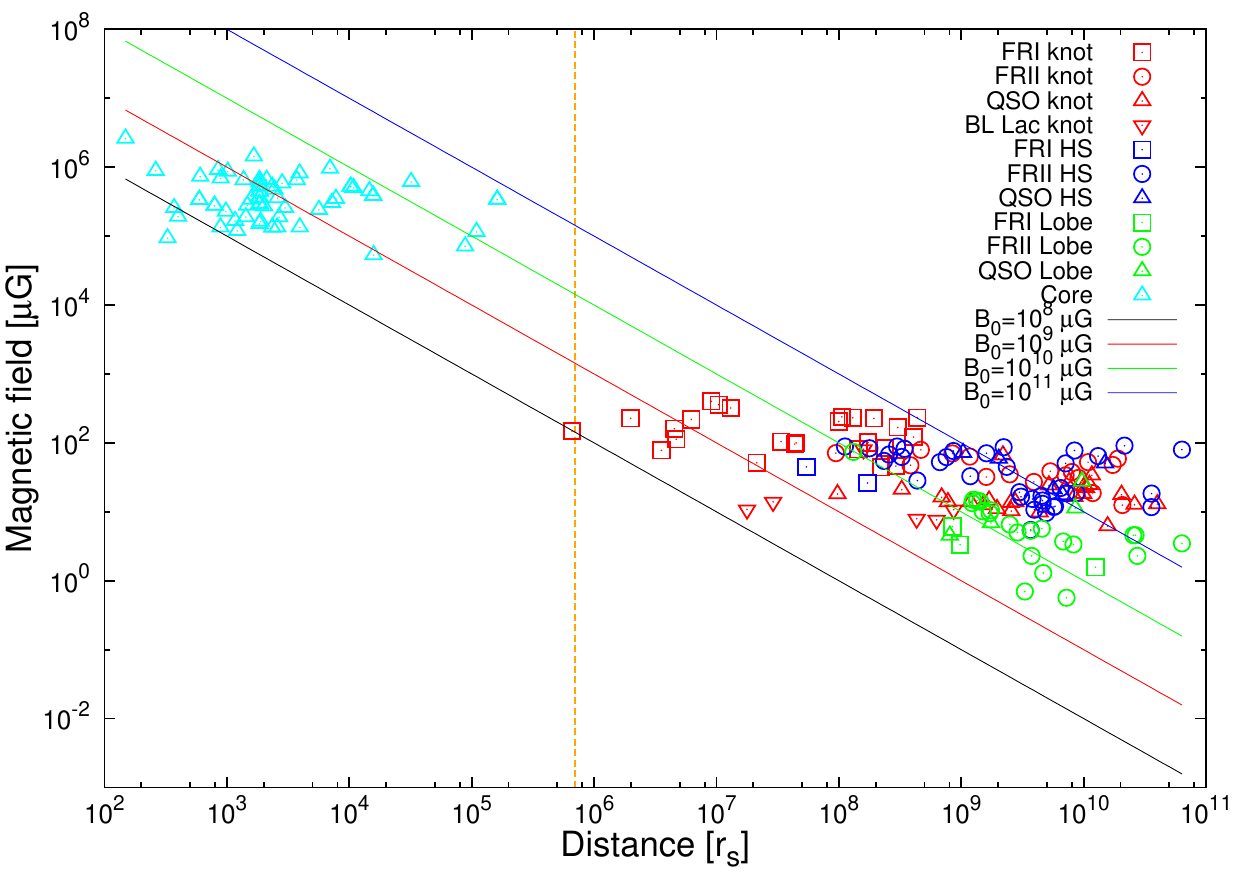}
        \caption{Observed equipartition magnetic field as a function of the distance from the jet base. The orange dashed line denotes the Bondi radius of AGN, whose gas temperature is 0.5 keV. The solid line represents the power-law function $B(d)=B_{\rm 0}(d/r_{\rm s})^{-1}$ where $B_{\rm 0}$ is the initial magnetic field at the Schwarzschild radius $r_{\rm s}$. }
        \label{fig:mag_distace}
    \end{center}
\end{figure*}
\begin{figure*}[t]
    \begin{center}
        \includegraphics[keepaspectratio,scale=1.3]{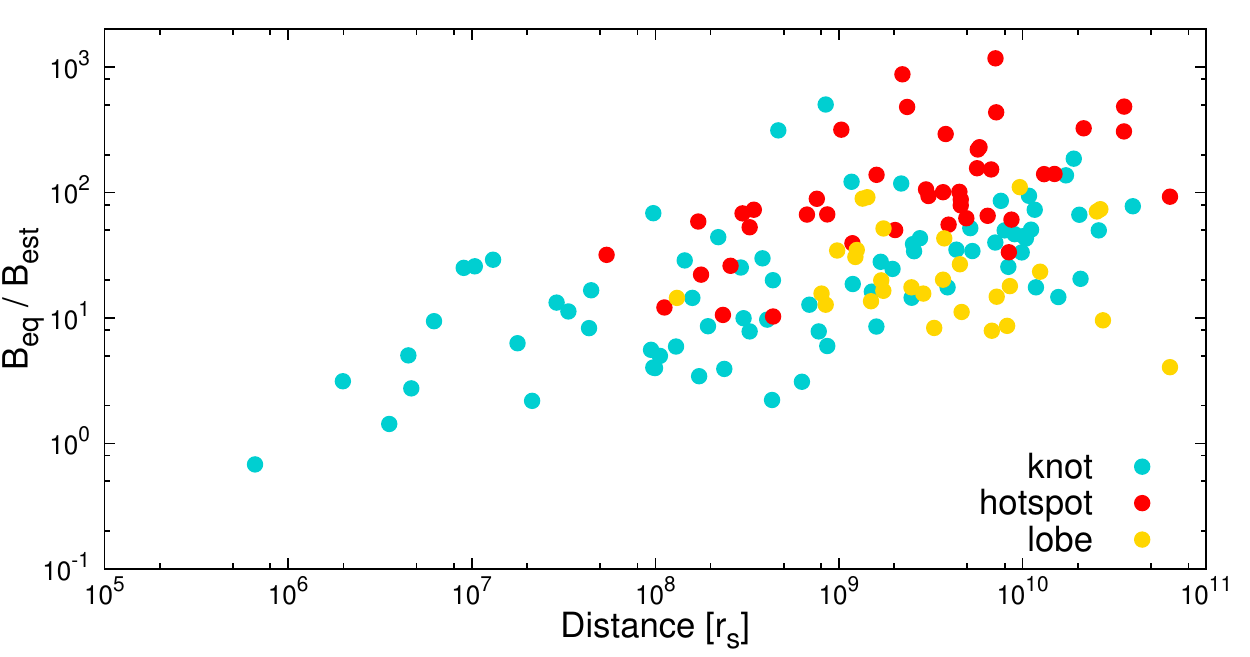}
        \caption{Ratio of the observed equipartition magnetic field $B_{\rm eq}$ to estimated magnetic field $B_{\rm est}$ plotted against distance in $r_{\rm s}$ from the jet base.}
        \label{fig:mag_ratio}
    \end{center}
\end{figure*}

To estimate the magnetic field strength, the X-ray emission mechanism must be determined using a detailed multi-wavelength spectral analysis \citep[see e.g.,][]{2002ApJ...564..683M}.
However, we can conduct such analysis for a few knots, although we can estimate the magnetic field for most lobes and even some hotspots from the X-ray through an inverse Compton emission \citep[see e.g.,][]{2005ApJ...626..733C}. Our purpose was to systematically investigate the spatial variation of magnetic fields along the AGN jets. Therefore, we approximate the magnetic field strengths in a homogeneous approach by assuming the energy equipartition between the electron and magnetic field energy density $u_{\rm e}, u_{\rm B}$ for all components.

Observationally, some of the jet components deviate from the equipartition \citep[e.g.,][]{2001ApJ...547..740W,2002ApJ...581..948H,2003A&A...399...91K}. For example,  \citet{2009ApJ...706..454I} showed that the upper limit of $u_{\rm e}/u_{\rm B}$ for radio lobes was $\sim$100. Here, the synchrotron luminosity scales with $u_{\rm e}u_{\rm B}V$, where $V$ is the volume of the emitting region. This means that $u_{\rm e} = 10u_{\rm e, eq}$ and $u_{\rm B} = 0.1u_{\rm B, eq}$ in this case, where $u_{\rm e, eq}$ and $u_{\rm B, eq}$ are $u_{\rm e}$ and $u_{\rm B}$ under the assumption of equipartition. Then, we may overestimate the magnetic field by approximately a factor of 3. However, as shown below, such uncertainty does not significantly change our main conclusions.

Under this assumption, we can derive the equipartition magnetic field $B_{\rm eq}$ from $V$ and the synchrotron luminosity $L_{\rm \nu}$ observed at frequency $\nu$. 
At $\delta=1$, $B_{\rm eq}$ can be represented by 

\begin{equation} \label{eq:B_eq}
    B_{\rm eq,\delta=1}\equiv\left[\frac{3\mu_{\rm 0}}{2}\frac{\eta\,G(\alpha_{\rm R},\nu,\nu_{\rm min}) L_{\rm \nu}}{V}\right]^{\frac{2}{7}}
\end{equation}
where $\mu_{\rm 0}$ is the permeability of a vacuum, $\eta$ is the ratio of the total energy density of protons and electrons to the energy density of electrons, and $G(\alpha_{\rm R},\nu,\nu_{\rm min})$ is a function given by \citet{1994hea2.book.....L}.
Here, $G(\alpha_{\rm R},\nu,\nu_{\rm min})$ depends slightly on $\alpha_{\rm R}$, $\nu$, and the minimum synchrotron frequency $\nu_{\rm min}$.

Because we use $L_{\nu}$ at 5~GHz, we further set $\nu_{\rm min}=5$~GHz following \citet{2005ApJ...622..797K}. Observationally, the positions of $\nu_{\rm min}$ vary with the sources, and are much lower frequencies of as low as $\sim$100~MHz \citep[see e.g.,][]{2020MNRAS.495..143C}. However, from Eq.~\ref{eq:B_eq}, the dependence on $\nu_{\rm min}$ of $B_{\rm eq, \delta=1}$ is $B_{\rm eq, \delta=1}\propto\nu_{\rm min}^{-(2 \alpha_{\rm R}-1)/7}$. As shown in Figure~\ref{fig:radio_index}, $\alpha_{\rm R}\le1.2$ in our jet component samples, which indicates $B_{\rm eq, \delta}\propto \nu_{\rm min}^{-1/5}$ in the strongest dependence case. Because $\nu_{\rm min}$ is as low as $\sim$100~MHz in AGN jets, the choice of $\nu_{\rm min}=5$~GHz in our study gives the lower limit estimate of $B_{\rm eq, \delta=1}$, but does not affect our results significantly because of its weak dependence on $\nu_{\rm min}$.

By setting $\nu=5\mathrm{GHz}$ and $\alpha_{\rm R}=0.75$, Eq \ref{eq:B_eq} can be written as

\begin{equation} \label{eq:B_eq2}
    \begin{split}
          B_{\rm eq,\delta=1}&\approx123\,\eta^{\frac{2}{7}}\left(1+z\right)^{\frac{11}{7}}\left(\frac{\theta}{0^{\prime\prime}3}\right)^{-\frac{6}{7}} \\
          &\quad\quad\times\left(\frac{d_{\rm L}}{100\,\mathrm{Mpc}}\right)^{-\frac{2}{7}}\left(\frac{f_{\rm 5}}{100\,\mathrm{mJy}}\right)^{\frac{2}{7}}\mathrm{\mu G}
    \end{split}
\end{equation}
(see, \cite{2005ApJ...622..797K}).
Finally, we correct $\delta$ to $B_{\rm eq,\delta=1}$ ,

\begin{equation} \label{eq:B_eq3}
  B_{\rm eq}=B_{\rm eq,\delta=1}\delta^{-\frac{5}{7}}
\end{equation}
\citep{2003ApJ...597..186S}.
Table \ref{tab:comp_data} lists $B_{\rm eq}$ for the cores and jet components.
We set $\eta=1$ in this paper, which means a pure leptonic jet. Because the jet composition is still under debate, the exact value of $\eta$ is uncertain. However, recent studies have suggested that a pure proton-electron jet is not energetically favored \citep{Sikora2020MNRAS.499.3749S}. Moreover, as seen in Eq~\ref{eq:B_eq2}, $B_{\rm eq}$ depends on $\eta^{2/7}$. Thus, this assumption will not significantly affect our results.

The derived $B_{\rm eq}$ is plotted against $d$ in $r_{\rm s}$ for each AGN in Figure~\ref{fig:mag_distace}.
Because we cannot obtain the black hole mass $M_{\rm BH}$ for 2 FR~Is and 12 FR~IIs, we set $M_{\rm BH}\sim 10^9 M_{\rm \odot}$, which is the average value of $M_{\rm BH}$ of the other AGNs used in this research. 
The tendency of $B_{\rm eq}$ to roughly decay inversely with $d$ is shown clearly in the figure.
A similar tendency has been previously reported.
\citet{2009ApJ...706..454I} indicated that $u_{\rm B}$ decays with $u_{\rm B} \propto d^{-2.4 \pm 0.4}$ for radio lobes within the range $d <$ 1 Mpc when using a magnetic field as determined through the inverse Compton X-ray emission.
In addition, \citet{2019ApJ...878..139T} revealed that the jet magnetic field $B$ decays with $B \propto d^{-0.88\pm0.14}$ by using the magnetic fields of seven FR I cores and two Cen A’s knots derived from the spectral energy distribution fit under the assumption of a synchrotron self-Compton model.
Our results suggest a similar trend not only for FR~Is but also for FR~IIs, QSOs, and BL Lacs.
In addition, it appears that the difference in the type of jet component rather than the type of AGN affects the spatial variations of the magnetic field along the jets.

Ideally, our investigation should consider the differences in the X-ray emission mechanisms. 
However, we cannot accurately estimate such an effect because there are few jet components whose X-ray emission mechanism is determined from the detailed multi-wavelength analysis in our sample. 
To evaluate this effect as much as possible, we compared our results with those of a multi-wavelength analysis \citep{2019ApJ...878..139T} by using the knots of Cen~A and found that they are roughly consistent (i.e., $B \sim 100 \mu$G at $d \sim 10^8 r_s$).
Therefore, the spatial variation of the jet magnetic field cannot be significantly changed, even if we consider a multi-wavelength analysis.

For comparison, Table \ref{tab:comp_data} lists the $estimated$ magnetic field strength $B_{\rm est}$ of jet components.
$B_{\rm est}$ are calculated from $B_{\rm eq}$ of the core, $B_{\rm eq,core}$, by assuming the dependence of $d^{-1}$, namely 

\begin{equation} \label{eq:B_estimate}
  B_{\rm est}=B_{\rm eq,core}\Bigl(\frac{d_{\rm core}}{d}\Bigr).
\end{equation}

Note that $B_{\rm est}$ is affected by the assumed $R$ as $B_{\rm eq,core}$, and $d_{\rm core}$ depends on $R$, that is, $B_{\rm eq,core}\propto\theta^{-6/7} \propto R^{-6/7}$ and $d_{\rm core} \propto R$. 
However, the total effect of $R$ on $B_{\rm est}$ is proportional to $R^{1/7}$, and the effect of the deviation between $R$ and the true core radius is small.


\section{Discussion} \label{sec:discussion}
\subsection{Properties of 57 AGNs}
Figure \ref{fig:mag_distace} shows a trend in which the magnetic field decreases along the jet axis, but rather gently around $10^{9-10}$ $r_{\rm s}$ away from the core. 
In addition, the magnetic field in different jet components appears to have a different dependence on $d$.
To clarify these trends, Figure~\ref{fig:mag_ratio} presents the ratio of $B_{\rm eq}$ to $B_{\rm est}$, i.e., $B_{\rm eq}/B_{\rm est}$, against $d$ in $r_{\rm s}$ of each AGN.
Clearly, $B_{\rm eq}$ of most jet components is significantly higher than $B_{\rm est}$.
Moreover, the deviation from $B_{\rm est}$ tends to increase as $d$ increases. 

The most likely explanation for this is the reduction in jet velocity. 
When the magnetic field energy is transported downstream of the jet as the Poynting flux, the magnetic field strength at distance $d$ is given as

\begin{equation} \label{eq:B_Poynt}
  B(d)= \frac{2}{\Gamma (d)
  R_{\rm jet}(d)}\sqrt{\frac{\xi_{\rm B} L_{\rm jet}}{v_{\rm jet}(d)}}\approx \frac{2}{d}\sqrt{\frac{\xi_{\rm B} L_{\rm jet}}{v_{\rm jet}(d)}},
 \end{equation}
where $\Gamma(d)$ is the bulk Lorentz factor at $d$, $R_{\rm jet}(d)$ is the jet radius at $d$, $\xi_{\rm B}$ is the energy fraction of the magnetic fields in the jet, $L_{\rm jet}$ is the total jet power, and $v_{\rm jet}(d)$ is the bulk speed of the jet at $d$.  

Considering the theoretically proposed jet launching mechanisms, $\xi_{\rm B}$ and $L_{\rm jet}$ cannot increase during its propagation \citep[see e.g.,][]{Sikora2020MNRAS.499.3749S}. Then, comparing Figure~\ref{fig:mag_distace} and Eq.~\ref{eq:B_Poynt}, the reduction of the jet velocities are the reason for the enhancement of the magnetic fields. If the energy injection in the jet or magnetic field is possible during the propagation, further complicated treatment may be needed, which is beyond the scope of this paper.

Figure \ref{fig:mag_ratio} also shows that the deviation of $B_{\rm eq}$ from $B_{\rm est}$ depends on the jet components. 
The geometric means of $B_{\rm eq}/B_{\rm est}$ is Hotspot : $97.1^{+37.8}_{-27.2}$  $>$ Lobe : $22.2^{+8.6}_{-6.2}$ $\geq$ Knot : $18.0^{+7.0}_{-5.1}$. 
The error indicates the range of $B_{\rm eq}/B_{\rm est}$ obtained if $R$ is assumed to range from $10^{16}$cm to $10^{18}$cm. 
These comparisons give similar results when we set $\delta = 1$ for hotspots, i.e., Hotspot : $114^{+44}_{-32}$ $>$ Lobe : $22.2^{+8.6}_{-6.2}$ $\geq$ Knot : $18.0^{+7.0}_{-5.1}$. 
Therefore, the magnetic fields of the hotspots were significantly amplified compared to the other jet components. 

\begin{figure}
    \centering
    \includegraphics[keepaspectratio,width=\linewidth]{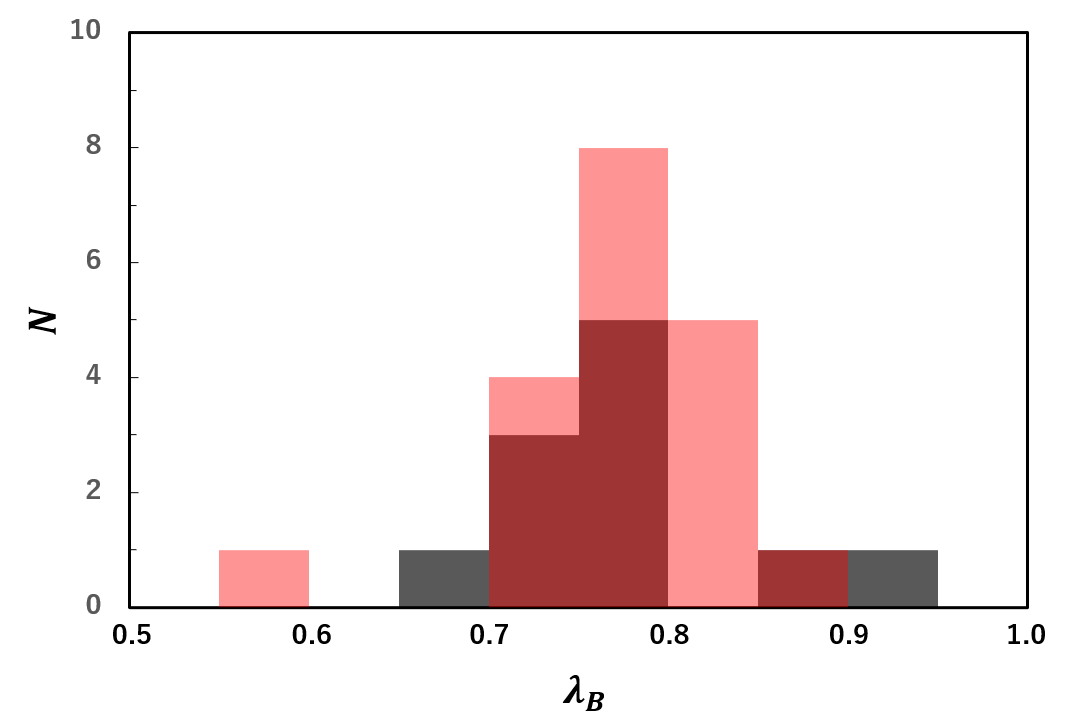}
    \caption{Distribution of the index $\lambda_{\rm B}$ obtained by fitting the magnetic field $B$ and distance from the jet base $d$ as $B\propto d^{-\lambda_{\rm B}}$ for FR Is(black) and FR IIs(red).}
    \label{fig:lambda_B}
\end{figure}

We further compared the spatial distribution of the magnetic fields of the FR~I and FR~II jets using  the index $\lambda_{\rm B}$ defined as $B\propto d^{-\lambda_{\rm B}}$. Because $B_{\rm eq}$ of the hotspots is amplified compared to that of the other jet components, we remove hotspots from the samples. Figure \ref{fig:lambda_B} shows the distribution of $\lambda_{\rm B}$ for FR~Is and FR~IIs. We do not see an apparent difference between FR~I and FR~II even though they have different jet powers and structures. Further investigation with larger samples is needed to investigate the underlying physics for this similarity.

\subsection{Magnetic Field at the Jet Base}
The initial magnetic fields at the jet bases at a scale of $\sim$10-100 $r_{\rm s}$ are roughly $\sim$10-100~G by simply extrapolating those at the core regions with $B\propto d^{-1}$, that is, a constant velocity is assumed. Observationally, it is not easy to explore the magnetic field at the jet base region. However, \citet{Kino2015ApJ...803...30K} determined the magnetic field of the M87 jet to be $\sim50$-124~G at a scale of $\sim3r_{\rm s}$ using the event horizon telescope data at 230~GHz based on the synchrotron self-absorption argument. \citet{Inoue2018ApJ...869..114I} determined the coronal magnetic field strength of nearby Seyferts using the cm-mm observations as $\sim10$~G at a scale of $\sim40r_{\rm s}$.  Therefore, these estimates are roughly consistent with each other. Although this similarity may come from various simplifications in our analysis, however, it may be interesting to investigate the origin of this similarity in future studies.

\subsection{Centaurus~A}
In this study, we discuss the spatial variation of AGN jet magnetic fields utilizing 57 AGNs. Ideally, such a study should be conducted independently for each source with a sufficient number of jet components. 
Hence, by focusing on Cen~A, where the inner jet was observed using TANAMI \citep{2014A&A...569A.115M}, we investigated the relationship between the magnetic field and the distance from the inner jet to the lobe located Mpc away from the core.

\begin{figure}
    \centering
    \includegraphics[keepaspectratio,width=\linewidth]{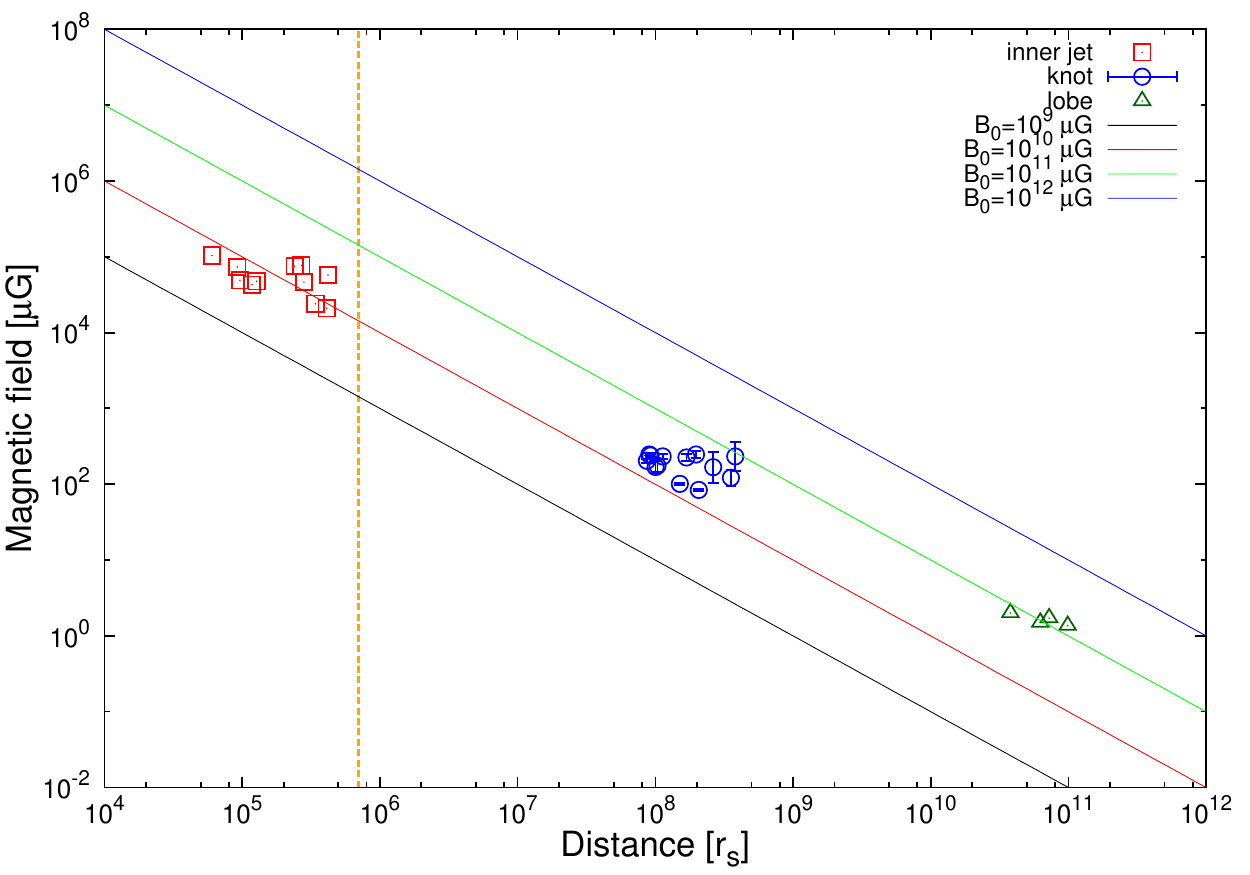}
    \caption{Observed equipartition magnetic field as a function of the distance from the jet base $d$ for Cen~A's emission region. The orange dashed line indicates the Bondi radius of Cen~A. The solid line shows the power-law function $B(d)=B_{\rm 0}(d/r_{\rm s})^{-1}$ where $B_{\rm 0}$ is the initial magnetic field at the Schwarzschild radius $r_{\rm s}$.}
    \label{fig:CenA_mag}
\end{figure}

Table \ref{tab:Cen_A_data} summarizes the properties of the knots and lobes of Cen~A. The inner jet uses knots with the FWHM exceeding the major radio beam size at least once in multiple observations. 
The volume of the lobe was calculated by assuming a cylindrical geometry.
Figure \ref{fig:CenA_mag} shows the magnetic field in Cen~A as a function of the distance from the core. 
The Bondi radius $r_{\rm B}$ shown in Figure \ref{fig:CenA_mag} is $7 \times 10^5 r_{\rm s}$, derived from $kT = 0.5$ keV \citep{2006PASJ...58..211H}, where $k$ is the Boltzmann constant, $T$ is the gas temperature, and the Schwarzschild radius of Cen~A is $r_{\rm s}\simeq 1.6\times10^{13}$cm \citep{2019ApJ...878..139T}.

\begin{figure}
    \centering
    \includegraphics[keepaspectratio,width=\linewidth]{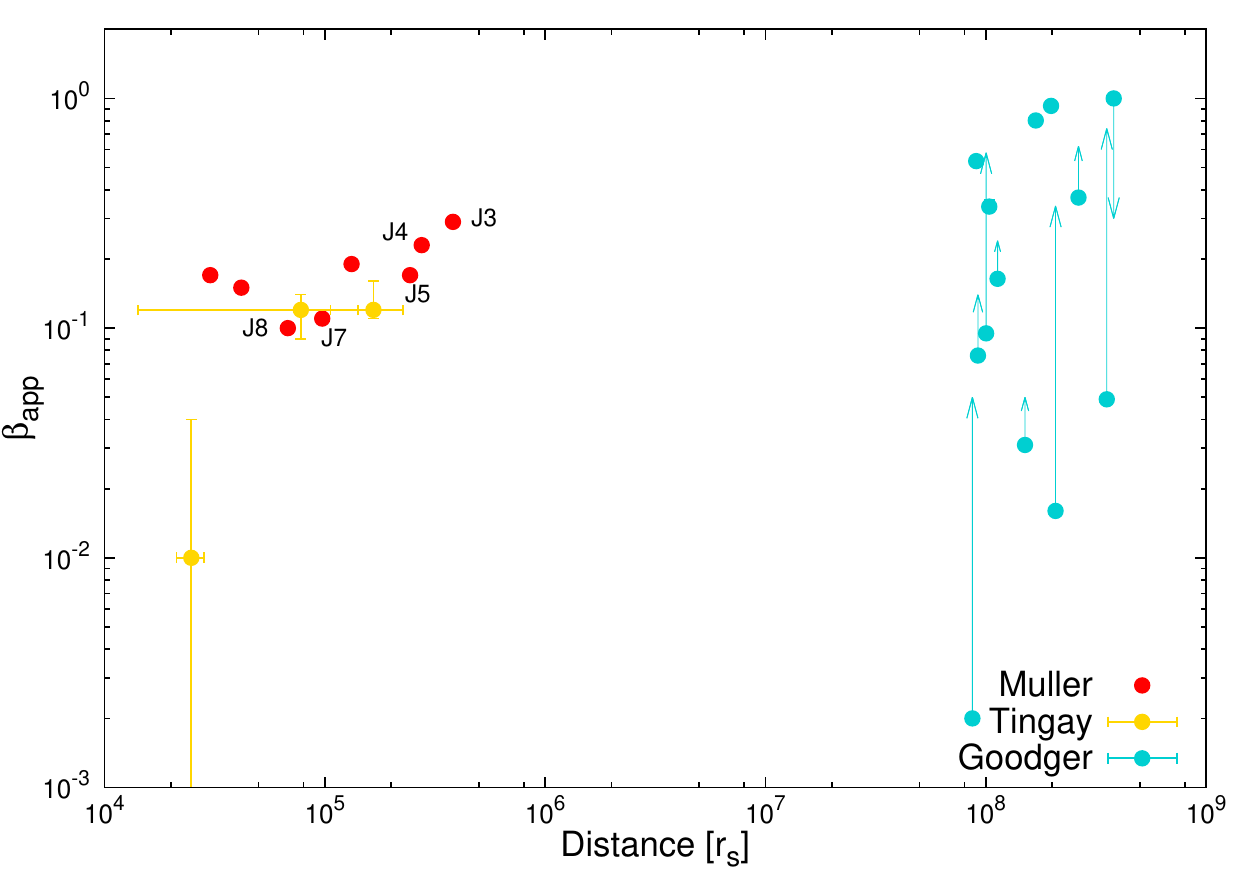}
    \caption{Observed velocity distribution for Cen A's knot located from upstream to downstream of the jet.}
    \label{fig:CenA_velocity}
\end{figure}

Figure \ref{fig:CenA_velocity} shows the velocity distributions of the jets observed by VLA, VLBA, and TANAMI \citep{2001AJ....122.1697T, 2010ApJ...708..675G, 2014A&A...569A.115M}. There are only three knots whose velocities were accurately measured at a few kpc away from the core, all of which had velocities of approximately $0.5c$. The jet velocity acceleration is observed in the inner jet, and reaches 0.5$c$ at 2.5 pc from the core \citep{2014A&A...569A.115M}.

\begin{figure}
    \centering
    \includegraphics[keepaspectratio,width=\linewidth]{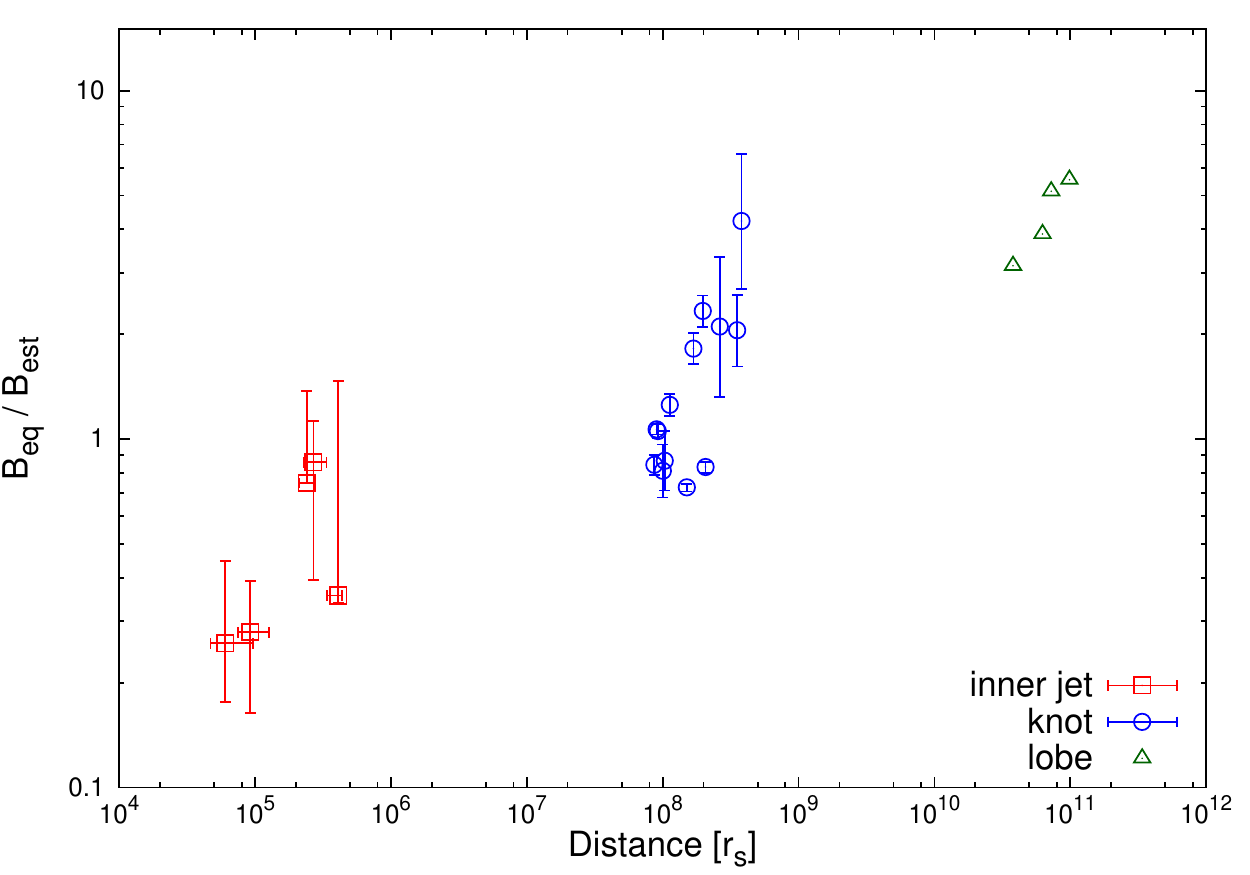}
    \caption{Ratio of the observed equipartition magnetic field $B_{\rm eq}$ to estimated magnetic field $B_{\rm est}$ plotted against distance from the core for Cen A's emission region. For the inner jet, $B_{\rm eq}/B_{\rm est}$ is only plotted when the observed knot's FWHM is larger than the radio beam size and the error indicates the range that $B_{\rm eq}/B_{\rm est}$ can take when the observed knot's FWHM is smaller than the radio beam size.}
    \label{fig:CenA_prop}
\end{figure}

Figure \ref{fig:CenA_prop} presents $B_{\rm eq}/B_{\rm est}$ against $d$ of Cen~A. We obtain $B_{\rm est}$ by using knot J2, whose FWHM is larger than the major beam size and whose velocity is considered to be close to that of the knots a few kpc away from the core (hereinafter referred to as ``kpc-scale knots"), as shown by $B_{\rm est}=B_{\rm J2}(d_{\rm J2}/{d})$, where $B_{\rm J2}$ is $B_{\rm eq}$ of knot J2 and $d_{\rm J2}$ is the distance from the core to knot J2.

The kpc-scale knots take $B_{\rm eq}/B_{\rm est}\simeq1$. 
Some of them show a magnetic field enhancement, which can be explained by a local amplification, such as a shock compression.
Such a magnetic amplification has been suggested based on the multi-wavelength data of Cen~A knots and diffuse jets \citep{2020ApJ...901L..27S}.
Because the magnetic field is not amplified owing to the jet deceleration at the kpc-scale knots, the jet velocity may not decay drastically from the inner jet to the kpc-scale knots.
On the other hand, the lobes take $B_{\rm eq}/B_{\rm est}\simeq 5$ and the magnetic field does not tend to decay as $d^{-1}$ at the terminus of the jet.
This tendency is consistent with the spatial variation of the 57 AGN jet magnetic fields.
Therefore, even if the discussion is based only on the observables, it can be confirmed that the deviation from $B_{\rm est}$ tends to be larger downstream where the jet decelerates.

\section{Conclusion} \label{sec:conclusion}
We systematically surveyed the spatial variation of the jet magnetic fields by utilizing cores, knots, hotspots, and lobes in 57 AGN samples.
We first calculated $B_{\rm eq}$ for each core and jet component. We also calculated $B_{\rm est}$ from $B_{\rm eq,core}$ for each jet component by assuming the relation $B \propto d^{-1}$.
Based on these calculations, we investigated the relationship between $B_{\rm eq}$ and $d$.
A resultant trend was revealed, in which the magnetic field decreases with increasing $d$. This can be confirmed in FR~Is, FR~IIs, QSOs, and BL~Lacs.
In addition, a comparison of $B_{\rm eq}/B_{\rm est}$ from upstream to downstream of the jet demonstrated that the magnetic fields were amplified owing to a jet deceleration.
Furthermore, a comparison of the geometric mean of $B_{\rm eq}/B_{\rm est}$ for jet components reveals that the magnetic fields of the hotspots are significantly amplified compared to the other jet components. 
Moreover, there is no clear difference in the amplification of the magnetic field owing to jet deceleration between FR~I and FR~II.
Finally, we investigated whether the spatial variation of the jet magnetic fields observed in multiple AGN jets holds even in a single jet by utilizing Cen~A. 
Consequently, the tendency of the magnetic field to decrease inversely with distance unless the jet decelerates is confirmed by a single jet even if the discussion is based only on the observables.

Many jet components discussed in this paper have X-ray counterparts, some of which have already been discussed in the literature \citep[e.g.,][]{2005ApJ...622..797K}. Implications on the variation of magnetic fields and jet velocity, as suggested in this paper, should be discussed within the framework of the spectral energy distribution including optical/X-ray counterparts, which remain as future work. Such direct comparison will provide further insight into the X-ray emission mechanism of large-scale jets and particle acceleration in relativistic jets in general.

\acknowledgements
Y.I. is supported by JSPS KAKENHI grant Nos. JP18H05458 and JP19K14772.
J.K. is supported by JSPS KAKENHI grant No. JP20K20923. We thanks an anonymous referee for his/her helpful comments and suggestions to improve the manuscript.

\begin{deluxetable*}{p{15em}cp{5em}p{8em}}
\tablenum{1}
\tablecaption{List of AGN with jet components\label{tab:core_data}}
\tablewidth{0pt}
\tablehead{
\colhead{Name} & \colhead{$\mathrm{Class^a}$} & \colhead{$z$}  & \colhead{Ref}
}
\decimalcolnumbers
\startdata
    3C 15............................................  &FR~I& 0.07338   &1,54 \\
    3C 31............................................  &FR~I& 0.0169    &2,3,23,55 \\
    3C 66B.........................................    &FR~I& 0.0215    &3,4 \\
    3C 346..........................................   &FR~I& 0.161     &3,5,56 \\
    3C 465...................................      &FR~I&0.0293&56,61\\
    4C +29.30..............................        &FR~I&0.064&38,62 \\
    M84..............................................  &FR~I& 0.00354   &6,7 \\
    M87..............................................  &FR~I& 0.00427   &8,9 \\ 
    Cen A...........................................   &FR~I& 0.00183   &10,11,12 \\ 
    Cen B....................................   &FR~I& 0.013 &45,63  \\ 
    NGC 315......................................      &FR~I& 0.0165    &3,13,45,55 \\
    Fornax A......................................     &FR~I& 0.00587   &14,45 \\
    3C 120..........................................   &FR~II& 0.033    &9,15,16,55 \\
    3C 403..........................................   &FR~II& 0.059    &17,18,46,54 \\
    3C 111..........................................   &FR~II& 0.0491   &19,70 \\
    Cygnus A.....................................      &FR~II& 0.0562   &20,45,46 \\
    Pictor A.......................................    &FR~II& 0.035    &10,21,45,47 \\
    3C 17.....................................  &FR~II&0.2197 &56,64 \\
    3C 33............................................  &FR~II& 0.0597   &22,46,55 \\
    3C 123..........................................   &FR~II& 0.218    &45,59,60\\
    3C 227...................................  &FR~II&0.0861 &56,65,66 \\
    3C 280..........................................   &FR~II& 0.996    &23,45,48 \\
    3C 294...................................  &FR~II&1.786 &18\\
    3C 295..........................................   &FR~II& 0.45     &24,45,48\\
    3C 303..........................................   &FR~II& 0.141    &3,47,49,56 \\
    3C 324...................................  &FR~II&1.2063 &18 \\
    3C 327...................................  &FR~II&0.1039  &65 \\
    3C 330..........................................   &FR~II& 0.55     &25,26,45 \\
    3C 390.3.......................................  &FR~II& 0.0561     &3,15,45,50,55 \\ 
    3C 445...................................  &FR~II&0.056  &56,65,67, \\
    PKS 2152$-$69........................  &FR~II&0.0283 &68 \\
    3C 6.1...........................................  &FR~II& 0.8404   &27,48 \\
    3C 47............................................  &FR~II& 0.425    &28,48,55 \\
    3C 109...........................................  &FR~II& 0.306     &3,23,29,48 \\
    3C 173.1.........................................  &FR~II& 0.292      &23,48,59 \\
    3C 219...........................................  &FR~II& 0.1744    &23,48,55 \\
    3C 265...........................................  &FR~II& 0.8108    &18,48 \\
    3C 321...........................................  &FR~II& 0.096     &23,48,58 \\
    3C 427.1.........................................  &FR~II& 0.572      &23,48\\
    3C 452...........................................  &FR~II& 0.0811    &3,18,23,48,51,58 \\
    4C 73.08.........................................  &FR~II& 0.0581     &30,45 \\
\enddata
\end{deluxetable*}

\begin{table*}
    \centering
	\caption{List of AGN with jet components}
	\begin{threeparttable}
		\begin{tabular}{p{15em}cp{5em}p{8em}}
		\hline \hline
		\multicolumn{1}{c}{Name} & $\mathrm{Class^a}$ & \multicolumn{1}{c}{$z$} & \multicolumn{1}{c}{Ref}\\
		\\
		\hline
    3C 179...........................................  &QSO(FSRQ)& 0.846       &31,32,45,55 \\
    3C 207...........................................  &QSO(SSRQ)& 0.684       &6,32,33,55 \\
    3C 345...........................................  &QSO(FSRQ)& 0.594       &6,9,34,55 \\
	4C 19.44.........................................  &QSO(FSRQ)& 0.72         &15,32,35,45,55\\
	PKS 1127$-$145..............................  &QSO(FSRQ)&  1.187   &15,36,57 \\
    3C 263...........................................          &QSO(SSRQ)&  0.656    &12,26,40,45,55 \\
    PKS 0637$-$752..............................  &QSO(FSRQ)& 0.653   &10,41,55 \\
    3C 273...........................................          &QSO(FSRQ)& 0.1583    &9,10,42,46,55 \\
    3C 380....................................  &QSO(SSRQ)&0.692 &55,69,70,71 \\
     3C 454.3.................................  & QSO(FSRQ)& 0.859 & 39,55,69,72 \\
     B2 0738+313..........................  & QSO(FSRQ)& 0.631 & 6,55,70,73 \\
     1928+738................................  & QSO(FSRQ)& 0.302 & 6,38,55,72 \\
    3C 371...........................................          &BL Lac& 0.051    &9,43,45,55,75 \\
    PKS 0521$-$365..............................  &BL Lac& 0.055    &15,44,55\\
     2201+044................................  & BL Lac& 0.027 & 55,74,75 \\
	\hline
	\end{tabular}
	\begin{tablenotes}
		\item NOTE---$\mathrm{^a}$\ FR~I: Fanaroff \& Riley class I radio galaxy;\ FR~II: Fanaroff \& Riley class II radio galaxy;\ QSO: quasar, either flat-spectrum radio quasars(FSRQ) or steep-spectrum radio quasars(SSRQ);\ BL Lac: BL Lac objects.\\
  1.\citet{2006ChJAA...6...25M},\citet{2003A&A...410..833K},\citet{1997MNRAS.291...20L}\ 2.\citet{2002MNRAS.334..182H},\citet{2002MNRAS.336..328L}\ 3.\citet{2001ApJ...552..508G}\ 4.\citet{2009ApJ...694.1485K},\citet{2001MNRAS.326.1499H},\citet{2003IAUJD..18E..39S}\ 5.\citet{2005MNRAS.360..926W},\citet{1995ApJ...452..605C},\citet{1994AJ....108..766B}\ 6.\citet{2004ApJ...616..110H}\ 7.\citet{2002ApJ...580..110H},\citet{2010ApJ...721..762W}\ 8.\citet{2001ApJ...559L..87N},\citet{2002ApJ...564..683M},\citet{1999ApJ...520..621B},\citet{2001ApJ...551..206P},\citet{2012ApJ...745L..28A}\ 9.\citet{1993ApJ...407...65G}\ 10.\citet{2003PASJ...55..351T}\ 11.\citet{2019ApJ...878..139T},\citet{2010ApJ...708..675G},\citet{2002ApJ...569...54K}\ 12.\citet{1994ApJ...430..467V}\ 13.\citet{1999ApJ...519..108C},\citet{2003MNRAS.343L..73W}\ 14.\citet{2020A&A...634A...9M},\citet{2001ApJ...546L..19T},\citet{2008MNRAS.391.1629N}\ 15.\citet{2008ApJS..175..314D}\ 16.\citet{2004ApJ...615..161H}\ 17.\citet{2012AIPC.1427..324T}\ 18.\citet{2004ApJ...612..729H}\ 19.\citet{2011ApJ...730...92H},\citet{2004ApJ...612..749Z},\citet{2016HEAD...1510606C},\citet{2011ApJ...734...43C}\ 20.\citet{1996ASPC..100..287C},\citet{2010ApJ...714...37Y},\citet{2004ApJ...609..539K},\citet{2007ApJ...662..213S},\citet{2003MNRAS.342..861T}\ 21.\citet{2016MNRAS.455.3526H},\citet{2006ApJ...642..711L}\ 22.\citet{2005ApJ...618..635G},\citet{2009A&A...498...61T},\citet{2007ApJ...659.1008K}\ 23.\citet{1988A&A...199...73G}\ 24.\citet{1991AJ....101.1623P},\citet{2001A&A...372..755B},\citet{2000ApJ...530L..81H} \ 25.\citet{1997AJ....114.2292F}\ 26.\citet{2002ApJ...581..948H}\ 27.\citet{1995ApJS...99..349N}\ 28.\citet{1994AJ....108..766B},\citet{1993ApJ...417..541V}\ 29.\citet{1994ApJ...435..116G},\citet{2006MNRAS.368.1395M}\ 30.\citet{1997A&A...328...78S}\ 31.\citet{2006AJ....131.1872P},\citet{2018ApJ...858...27Z}\ 32.\citet{2002ApJ...571..206S}\ 33.\citet{2002AJ....123.1258H},\citet{2017RAA....17...90X},\citet{2002A&A...381..795B}\ 34.\citet{2005AJ....130.1418J},\citet{2012ApJ...748...81K}\ 35.\citet{2017ApJ...846..119H}\ 36.\citet{2007ApJ...657..145S},\citet{2002ApJ...570..543S},\citet{2012AASP....2...49M}\ 37.\citet{2007ApJ...661..719U}\ 38.\citet{2004ApJ...608..698S}\ 39.\citet{2007ApJ...662..900T}\ 40.\citet{2006MNRAS.372..113M}\ 41.\citet{2000ApJ...540L..69S}\ 42.\citet{2016ApJ...818..195M},\citet{2007MNRAS.380..828J}, \citet{2020MNRAS.497.2066F}\ 43.\citet{2012ApJ...744..177L},\citet{2011MNRAS.414.2674G}\ 44.\citet{2016A&A...586A..70L},\citet{2002MNRAS.335..142B},\citet{1999ApJ...526..643S}\ 45.\citet{2005ApJ...622..797K}\ 46.\citet{2018ApJ...858...27Z}\ 47.\citet{1997A&A...325...57M}\ 48.\citet{2005ApJ...626..733C}\ 49.\citet{2003A&A...399...91K}\ 50.\citet{1998ApJ...499L.149H}\ 51.\citet{2002ApJ...580L.111I}\ 52.\citet{2006ApJ...641..717S}\ 53.\citet{2019APh...107...15M}\ 54.\citet{2014MNRAS.440..269M}\ 55.\citet{2002ApJ...579..530W}\ 56.\citet{2007ApJ...658..815S}\ 57.\citet{2004ApJ...602..103F}\ 58.\citet{2009A&A...495.1033B}\ 59.\citet{2018RAA....18..108Y}\ 60.\citet{2000ApJ...534..172L}\ 61.\citet{2005MNRAS.359.1007H},\citet{2009ApJ...694..992L},\citet{2020MNRAS.496..676B}\ 62.\citet{2004ApJ...608..698S},\citet{2009A&A...505..509L}\ 63.\citet{2008MNRAS.384..775M},\citet{2019A&A...627A.148A}\ 64.\citet{2000A&A...363...84V},\citet{2009ApJ...696..980M}\ 65.\citet{2008ApJ...678..712D}\ 66.\citet{2007ApJ...669..893H}\ 67.\citet{2011MNRAS.414.2739B},\citet{2012MNRAS.419.2338O}\ 68.\citet{2006MNRAS.371..898S},\citet{2012MNRAS.424.1346W}\ 69.\citet{1989AJ.....98.1208H}\ 70.\citet{2009ApJ...706.1253H}\ 71.\citet{2013PASJ...65...29K}\ 72.\citet{1999ApJ...521..493L}\ 73.\citet{2003ApJ...595..643S}\ 74.\citet{2004ApJ...613..752G},\citet{2009ASPC..402..233S}\ 75.\citet{2007ApJ...670...74S}
	\end{tablenotes}
	\end{threeparttable}
\end{table*}

\begin{deluxetable*}{ccccccccccc}
\tablenum{2}
\tablecaption{Parameters for Knot, Hotspot, Lobe \label{tab:comp_data}}
\tablewidth{0pt}
\tablehead{
\colhead{Name} & \colhead{Component$\mathrm{^a}$} & \colhead{$\alpha_{\rm R}$}  & \colhead{$f_{\rm 5}$} &
\colhead{$\theta$} & \colhead{$\delta$} & \colhead{$\theta_{\rm D}$} & \colhead{$\theta_{\rm inc}$} & \colhead{$\log \frac{M_{\rm BH}}{M_{\odot}}$}  & \colhead{$B_{\rm eq}$}  & \colhead{$B_{\rm est}$}  \\
\nocolhead{Common} & \nocolhead{Common} & \nocolhead{Common}  & \colhead{[mJy]} &
\colhead{[arcsec]} & \nocolhead{Common} & \colhead{[arcsec]} & \nocolhead{Common} & \nocolhead{Common} & \colhead{[$\mu$G]}  & \colhead{[$\mu$G]}  
}
\decimalcolnumbers
\startdata
3C 15&Core&&64&2.36E$-$05&1.03&&50&8.7&2.77E+05&\\
&K-C&0.9&55&$0.3\mathrm{^b}$&1.03&4.1&&&81&2.81\\
&L&0.75&1.80E+03&20&$1\mathrm{^b}$&25&&&6.1&0.48\\
3C 31&Core&&92&9.69E$-$05&1.0&&52&8.5&1.33E+05&\\
&K&0.55&37&$0.3\mathrm{^b}$&1.0&2.4&&&104&9.25\\
3C 66B&Core&&186&7.59E$-$05&1.0&&45&8.6&1.89E+05&\\
&K-A&0.75&3.9&$0.3\mathrm{^b}$&1.0&1.3&&&52&24\\
&K-B&0.6&34&$0.3\mathrm{^b}$&1.0&2.6&&&96&12\\
3C 346&Core&&329&1.18E$-$05&2.92&&20&8.8&3.39E+05&\\
&K-C&&158&$0.3\mathrm{^b}$&2.92&2.2&&&46&1.82\\
 3C 465& Core&& 267&5.60E$-$05&1.15&&60& 8.6&2.26E+05&\\
&K-B&&9.7&$0.3\mathrm{^b}$&1.15&5.6&&&44&1.01\\
4C +29.30&Core&&63&2.67E$-$05&1.55&&40&$9.0\mathrm{^b}$&1.91E+05&\\
&HS-B&&18&$0.3\mathrm{^b}$&3&18&&&45&1.41\\
&HS-C&&14&$0.3\mathrm{^b}$&3&11.1&&&26&0.45\\
M84& Core &&3.60E+03&3.96E$-$04&1.02&&45&8.9&1.66E+05&\\
&K-2.5&0.65&3.5&$0.3\mathrm{^b}$&1.02&2.5&&&78&54\\
&K-3.3&0.65&13&$0.3\mathrm{^b}$&1.02&3.3&&&113&41\\
M87&Core&&3.10E+03&4.22E$-$04&1.45&&20&9.5&1.19E+05&\\
&HST-1&0.71&77&$0.3\mathrm{^b}$&1.45&0.9&&&149&219\\
&K-D&0.69&337&$0.3\mathrm{^b}$&1.45&2.7&&&227&73\\
&K-E&0.71&101&$0.3\mathrm{^b}$&1.45&6.2&&&161&32\\
&K-F&0.69&296&$0.3\mathrm{^b}$&1.45&8.5&&&218&23\\
&K-A&0.67&2.46E+03&$0.3\mathrm{^b}$&1.45&12.3&&&400&16\\
&K-B&0.67&1.67E+03&$0.3\mathrm{^b}$&1.45&14.2&&&358&14\\
&K-C&0.69&1.15E+03&$0.3\mathrm{^b}$&1.45&17.8&&&322&11\\
Cen A&Core&&5.88E+03&1.77E$-$03&1.22&&30&7.7&7.10E+04&\\
&AX1A&&36&$0.3\mathrm{^b}$&1.22&14.1&&&204&51\\
&AX1C&&64&$0.3\mathrm{^b}$&1.22&15.0&&&239&48\\
&AX2&&58&$0.3\mathrm{^b}$&1.22&18.3&&&232&39\\
&AX3&&3.1&$0.3\mathrm{^b}$&1.22&24.5&&&101&29\\
&AX4&&52&$0.3\mathrm{^b}$&1.22&27.4&&&226&26\\
&AX5&&1.6&$0.3\mathrm{^b}$&1.22&33.6&&&84&21\\
&AX6&&18&$0.3\mathrm{^b}$&1.22&42.7&&&167&17\\
&BX2&&5.9&$0.3\mathrm{^b}$&1.22&57.5&&&121&13\\
&BX4&&58&$0.3\mathrm{^b}$&1.22&61.8&&&232&12\\
Cen B&Core&&6.58E+03&1.24E$-$04&1.81&&$30\mathrm{^b}$&$9.0\mathrm{^b}$&2.56E+05&\\
&Lobe&&3.70E+04&180&$1\mathrm{^b}$&229&&&3.3&0.10\\
NGC 315&Core&&689&9.79E$-$05&1.37&&35&8.9&1.88E+05&\\
&K&&68&$0.3\mathrm{^b}$&1.37&5.9&&&100&6.02\\
\enddata
\end{deluxetable*}

\begin{deluxetable*}{ccccccccccc}
\tablenum{2}
\tablecaption{Continued}
\tablewidth{0pt}
\tablehead{
\colhead{Name} & \colhead{Component$\mathrm{^a}$} & \colhead{$\alpha_{\rm R}$}  & \colhead{$f_{\rm 5}$} &
\colhead{$\theta$} & \colhead{$\delta$} & \colhead{$\theta_{\rm D}$} & \colhead{$\theta_{\rm inc}$} & \colhead{$\log \frac{M_{\rm BH}}{M_{\odot}}$}  & \colhead{$B_{\rm eq}$}  & \colhead{$B_{\rm est}$}  \\
\nocolhead{Common} & \nocolhead{Common} & \nocolhead{Common}  & \colhead{[mJy]} &
\colhead{[arcsec]} & \nocolhead{Common} & \colhead{[arcsec]} & \nocolhead{Common} & \nocolhead{Common} & \colhead{[$\mu$G]}  & \colhead{[$\mu$G]}  
}
\decimalcolnumbers
\startdata
Fornax A&Core&&43&3.25E$-$04&$1\mathrm{^a}$&&$30\mathrm{^b}$&8.2&5.30E+04&\\
&L&0.9&1.60E+04&450&$1\mathrm{^b}$&900&&&1.6&0.07\\
3C 120&Core&&5.54E+03&4.99E$-$05&10.98&&5&7.4&1.15E+05&\\
&K-4&0.74&66&$0.3\mathrm{^b}$&10.98&4&&&19&1.07\\
&K-7&0.68&16&$0.3\mathrm{^b}$&10.98&7&&&13&0.61\\
3C 403&Core&&7.1&2.88E$-$05&$1\mathrm{^a}$&&50&9.0&1.34E+05&\\
&K-F1&&27&$0.3\mathrm{^b}$&1&50.7&&&70&0.14\\
&K-F6&&41&$0.3\mathrm{^b}$&1&28.0&&&79&0.25\\
3C111&Core&&2.82E+03&3.42E$-$05&2.15&&18&8.3&3.84E+05&\\
&K-K9&&10&$0.3\mathrm{^b}$&2.15&9&&&32&3.77\\
&K-K14&&14&$0.3\mathrm{^b}$&2.15&14&&&35&2.43\\
&K-K22&&5.6&$0.3\mathrm{^b}$&2.15&22&&&27&1.54\\
&K-K30&&19&$0.3\mathrm{^b}$&2.15&30&&&39&1.13\\
&K-K40&&12&$0.3\mathrm{^b}$&2.15&40&&&34&0.85\\
&K-K45&&18&$0.3\mathrm{^b}$&2.15&45&&&38&0.75\\
&K-K51&&8.9&$0.3\mathrm{^b}$&2.15&51&&&31&0.67\\
&K-K61&&56&$0.3\mathrm{^b}$&2.15&61&&&52&0.56\\
&K-K97&&41&$0.3\mathrm{^b}$&2.15&97&&&48&0.35\\
&K-K107&&86&$0.3\mathrm{^b}$&2.15&107&&&59&0.32\\
&H-N&&901&$0.3\mathrm{^b}$&$3\mathrm{^b}$&121&&&91&0.28\\
&H-S&&269&$0.3\mathrm{^b}$&$3\mathrm{^b}$&74&&&65&0.46\\
Cygnus A&Core&&3.66E+05&3.02E$-$05&1.15&&60&9.4&2.61E+06&\\
&HS-A&0.55&4.46E+04&1.2&$3\mathrm{^b}$&66.5&&&82&1.1\\
&HS-D&0.5&5.77E+04&1.2&$3\mathrm{^b}$&57.7&&&89&1.3\\
Pictor A&Core&&1.24E+03&4.71E$-$05&1.4&&45&7.6&3.39E+05&\\
&HS&0.74&2000&0.5&$3\mathrm{^b}$&250&&&80&0.86\\
&L&0.72&13000&90&$1\mathrm{^b}$&250&&&3.5&0.86\\
3C 17&Core&&316&9.21E$-$06&1.41&&45& 8.7&6.85E+05&\\
&K-S3.7&&30&0.3&1.41&3.7&&&47&1.59\\
&K-S11.3&&83&0.3&1.41&11.3&&&63&0.52\\
3C 33&Core&&10.6&2.84E$-$05&0.9&&75&8.4&1.62E+05&\\
&HS-S1&0.75&613&0.5&$3\mathrm{^b}$&138.7&&&50&0.04\\
&HS-S2&0.98&498&1.5&$3\mathrm{^b}$&140&&&19&0.04\\
&HS-N1&0.88&63&1.25&$3\mathrm{^b}$&113.7&&&12&0.05\\
&HS-N2&0.9&59&1.25&$3\mathrm{^b}$&111.2&&&12&0.05\\
3C 123&Core&&93&9.29E$-$06&$1\mathrm{^b}$&&$30\mathrm{^b}$&7.9&6.12E+05&\\
&HS&0.5&5.20E+03&0.5&$3\mathrm{^b}$&8.6&&&78&2.32\\
3C 227&Core&&32&2.03E$-$05&$1\mathrm{^a}$&&$30\mathrm{^b}$&8.9&2.56E+05&\\
&HS-P1&&24&0.66&$3\mathrm{^b}$&110&&&15&0.17\\
&HS-P2&&37&0.87&$3\mathrm{^b}$&110&&&13&0.17\\
&HS-P3&&19&1   &$3\mathrm{^b}$&117.8&&&9.6&0.15\\
\enddata
\end{deluxetable*}

\begin{deluxetable*}{ccccccccccc}
\tablenum{2}
\tablecaption{Continued}
\tablewidth{0pt}
\tablehead{
\colhead{Name} & \colhead{Component$\mathrm{^a}$} & \colhead{$\alpha_{\rm R}$}  & \colhead{$f_{\rm 5}$} &
\colhead{$\theta$} & \colhead{$\delta$} & \colhead{$\theta_{\rm D}$} & \colhead{$\theta_{\rm inc}$} & \colhead{$\log \frac{M_{\rm BH}}{M_{\odot}}$}  & \colhead{$B_{\rm eq}$}  & \colhead{$B_{\rm est}$}  \\
\nocolhead{Common} & \nocolhead{Common} & \nocolhead{Common}  & \colhead{[mJy]} &
\colhead{[arcsec]} & \nocolhead{Common} & \colhead{[arcsec]} & \nocolhead{Common} & \nocolhead{Common} & \colhead{[$\mu$G]}  & \colhead{[$\mu$G]}  
}
\decimalcolnumbers
\startdata
3C 227&HS-P4&&1.8&0.4&$3\mathrm{^b}$&94.5&&&11&0.19\\
&HS-F1&&7.8&0.38&$3\mathrm{^b}$&108&&&17&0.17\\
3C 280&Core&&1.0&4.03E$-$06&$1\mathrm{^a}$&&$30\mathrm{^b}$&9.0&4.42E+05&\\
&HS-W&0.8&720&$0.3\mathrm{^b}$&$3\mathrm{^b}$&0.86&&&88&7.3\\
&HS-E&0.8&330&$0.3\mathrm{^b}$&$3\mathrm{^b}$&12.2&&&71&0.51\\
&L-W&&39&3.8&$1\mathrm{^b}$&13&&&9.6&0.48\\
3C 294&Core&&0.53&3.80E$-$06&$1\mathrm{^a}$&&$30\mathrm{^b}$&$9.0\mathrm{^b}$&5.33E+05&\\
&HS-N&&143&0.3&$3\mathrm{^b}$&6.2&&&76&1.14\\
&HS-S&&22&0.43&$3\mathrm{^b}$&8.5&&&33&0.83\\
3C 295&Core&&3.0&5.66E$-$06&$1\mathrm{^a}$&&$30\mathrm{^b}$&9.0&3.63E+05&\\
&HS-NW&0.65&1.30E+03&$0.3\mathrm{^b}$&$3\mathrm{^b}$&1.9&&&84&3.78\\
&HS-SE&0.65&630&$0.3\mathrm{^b}$&$3\mathrm{^b}$&2.8&&&68&2.61\\
&L&0.9&6.50E+03&1.5&$1\mathrm{^b}$&1.4&&&74&5.1\\
3C 303&Core&&181&1.32E$-$05&1.98&&30&8.0&3.48E+05&\\
&HS&0.84&260&1.0&$3\mathrm{^b}$&17&&&19&0.31\\
3C 324&Core&&0.14&3.88E$-$06&$1\mathrm{^a}$&&$30\mathrm{^b}$&$9.0\mathrm{^b}$&2.85E+05&\\
&HS-E&&277&0.365&$3\mathrm{^b}$&5.57&&&62&0.70\\
&HS-W&&85&0.3&$3\mathrm{^b}$&4.92&&&52&0.79\\
3C 327&Core&&34&1.72E$-$05&$1\mathrm{^a}$&&$30\mathrm{^b}$&$9.0\mathrm{^b}$&2.94E+05&\\
&HS-S1&&22&0.45&$3\mathrm{^b}$&97&&&19&0.18\\
&HS-S2&&4.7&0.32&$3\mathrm{^b}$&100&&&17&0.18\\
3C 330&Core&&0.74&5.06E$-$06&$1\mathrm{^a}$&&45&9.0&2.77E+05&\\
&HS-NE&1.0&1.30E+03&$0.3\mathrm{^b}$&$3\mathrm{^b}$&30.1&&&87&0.10\\
&HS-SW&1.0&130&$0.3\mathrm{^b}$&$3\mathrm{^b}$&31.9&&&45&0.09\\
&L-NE&0.9&260&3.5&$1\mathrm{^b}$&18.2&&&15&0.16\\
&L-SW&1.0&230&3.5&$1\mathrm{^b}$&19.4&&&14&0.15\\
3C 390.3&Core&&120&3.02E$-$05&0.94&&30&8.6&3.05E+05&\\
&HS-NE-B&0.7&350&1.1&$3\mathrm{^b}$&102.3&&&22&0.34\\
3C 445&Core&&62&3.02E$-$05&$1\mathrm{^a}$&&60&8.0&2.41E+05&\\
&HS-SE&&98&1.52&$3\mathrm{^b}$&276.1&&&12&0.04\\
&HS-SW&&51&0.72&$3\mathrm{^b}$&276.6&&&18&0.04\\
PKS 2152-69&Core&&8.30E+03&5.80E$-$05&$1\mathrm{^a}$&&$30\mathrm{^b}$&$9.0\mathrm{^b}$&6.55E+05&\\
&HS-N&0.7&63&0.56&$3\mathrm{^b}$&48.0 &&&28&2.77\\
&HS-S&0.7&167&0.36&$3\mathrm{^b}$&25.6&&&55&5.18\\
&K-D&0.75&15&0.3&$1\mathrm{^a}$&10.4&&&71&12.8\\
3C 6.1&Core&&4.4&4.24E$-$06&$1\mathrm{^a}$&&$30\mathrm{^b}$&9.0&6.05E+05&\\
&L-N&&156&4.96&$1\mathrm{^b}$&14&&&11&0.64\\
&L-S&&84&4.22&$1\mathrm{^b}$&12&&&10&0.75\\
3C 47&Core&&74&5.85E$-$06&1.1&&30&8.7&8.13E+05&\\
&L-N&&198&15&$1\mathrm{^b}$&31.6&&&3.7&0.47\\
&L-S&&238&17.96&$1\mathrm{^b}$&38.2&&&3.4&0.39\\
\enddata
\end{deluxetable*}

\begin{deluxetable*}{ccccccccccc}
\tablenum{2}
\tablecaption{Continued}
\tablewidth{0pt}
\tablehead{
\colhead{Name} & \colhead{Component$\mathrm{^a}$} & \colhead{$\alpha_{\rm R}$}  & \colhead{$f_{\rm 5}$} &
\colhead{$\theta$} & \colhead{$\delta$} & \colhead{$\theta_{\rm D}$} & \colhead{$\theta_{\rm inc}$} & \colhead{$\log \frac{M_{\rm BH}}{M_{\odot}}$}  & \colhead{$B_{\rm eq}$}  & \colhead{$B_{\rm est}$}  \\
\nocolhead{Common} & \nocolhead{Common} & \nocolhead{Common}  & \colhead{[mJy]} &
\colhead{[arcsec]} & \nocolhead{Common} & \colhead{[arcsec]} & \nocolhead{Common} & \nocolhead{Common} & \colhead{[$\mu$G]}  & \colhead{[$\mu$G]}  
}
\decimalcolnumbers
\startdata
3C 109&Core&&263&7.25E$-$06&2.18&&25&8.3&5.83E+05&\\
&L-N&&291&13&$1\mathrm{^b}$&45.9&&&4.6&0.06\\
&L-S&&332&13.6&$1\mathrm{^b}$&47.9&&&4.6&0.06\\
3C 173.1&Core&&7.4&7.47E$-$06&$1\mathrm{^a}$&&$30\mathrm{^b}$&9.0&3.57E+05&\\
&L-N&&271&8.34&$1\mathrm{^b}$&25&&&6.5&0.37\\
&L-S&&215&10.5&$1\mathrm{^b}$&29&&&5.0&0.32\\
3C 219&Core&&51&1.11E$-$05&$1\mathrm{^a}$&&$30\mathrm{^b}$&8.2&4.49E+05&\\
&L-N&0.8&1.03E+03&45.3&$1\mathrm{^b}$&73&&&2.3&0.24\\
3C 265&Core&&2.89&4.29E$-$06&$1\mathrm{^a}$&&$30\mathrm{^b}$&9.0&5.24E+05&\\
&L-E&&97&9.3&$1\mathrm{^b}$&30&&&5.3&0.26\\
&L-W&&77&7.9&$1\mathrm{^b}$&37&&&5.7&0.21\\
3C 321&Core&&30&1.85E$-$05&$1\mathrm{^a}$&&$30\mathrm{^b}$&9.1&2.69E+05&\\
&L&&42&33.1&$1\mathrm{^b}$&150&&&1.3&0.12\\
3C 427.1&Core&&0.8&4.98E$-$06&$1\mathrm{^a}$&&$30\mathrm{^b}$&9.0&2.90E+05&\\
&L-N&&495&4.85&$1\mathrm{^b}$&11.6&&&13&0.44\\
&L-S&&611&4.59&$1\mathrm{^b}$&11.8&&&15&0.43\\
3C 452&Core&&130&2.15E$-$05&1.15&&60&8.8&3.36E+05&\\
&HS-W&&33&0.705&$3\mathrm{^b}$&130&&&15&0.05\\
&HS-E&&67&3&$3\mathrm{^b}$&126&&&5&0.05\\
&L&0.78&4.00E+03&80&$1\mathrm{^b}$&128&&&2.3&0.05\\
4C 73.08&Core&&11&2.92E$-$05&$1\mathrm{^a}$&&$30\mathrm{^b}$&9.0&1.50E+05&\\
&L-E&0.85&270&180&$1\mathrm{^b}$&400&&&0.57&0.04\\
&L-W&0.85&560&180&$1\mathrm{^b}$&183&&&0.7&0.08\\
3C 179&Core&&646&4.23E$-$06&6.72&&8.1&8.7&6.48E+05&\\
&K-A&0.8&73&$0.3\mathrm{^b}$&6.72&4.3&&&24&0.46\\
&K-B&0.8&110&$0.3\mathrm{^b}$&6.72&6.3&&&27&0.32\\
&CL&&290&2.0&$1\mathrm{^b}$&8&&&28&0.24\\
3C 207&Core&&1.30E+03&4.62E$-$06&9.89&&5.8&8.5&5.21E+05&\\
&K&0.8&230&$0.3\mathrm{^b}$&9.89&4.6&&&24&0.53\\
&HS&0.8&160&$0.3\mathrm{^b}$&$3\mathrm{^b}$&6.8&&&50&0.36\\
&L&0.9&250&5.0&$1\mathrm{^b}$&3.9&&&11&0.63\\
3C 345&Core&&8.40E+03&4.88E$-$06&21.1&&2.7&9.4&4.74E+05&\\
&K-A&&181.63&$0.3\mathrm{^b}$&21.1&2.7&&&12&0.77\\
4C 19.44&Core&&2.70E+03&4.49E$-$06&6.91&&8.2&9.4&8.62E+05&\\
&K-A&&57&$0.3\mathrm{^b}$&6.91&1.7&&&21&2.69\\
&K-B&&23&$0.3\mathrm{^b}$&6.91&3.6&&&16&1.27\\
&K-C&&13&$0.3\mathrm{^b}$&6.91&6.2&&&14&0.74\\
&K-D&&16&$0.3\mathrm{^b}$&6.91&8.8&&&15&0.52\\
&K-E&&6&$0.3\mathrm{^b}$&6.91&10.2&&&11&0.45\\
&K-F&&12&$0.3\mathrm{^b}$&6.91&13.2&&&13&0.35\\
&K-G&&13&$0.3\mathrm{^b}$&6.91&14.4&&&14&0.32\\
\enddata
\end{deluxetable*}

\begin{deluxetable*}{ccccccccccc}
\tablenum{2}
\tablecaption{Continued}
\tablewidth{0pt}
\tablehead{
\colhead{Name} & \colhead{Component$\mathrm{^a}$} & \colhead{$\alpha_{\rm R}$}  & \colhead{$f_{\rm 5}$} &
\colhead{$\theta$} & \colhead{$\delta$} & \colhead{$\theta_{\rm D}$} & \colhead{$\theta_{\rm inc}$} & \colhead{$\log \frac{M_{\rm BH}}{M_{\odot}}$}  & \colhead{$B_{\rm eq}$}  & \colhead{$B_{\rm est}$}  \\
\nocolhead{Common} & \nocolhead{Common} & \nocolhead{Common}  & \colhead{[mJy]} &
\colhead{[arcsec]} & \nocolhead{Common} & \colhead{[arcsec]} & \nocolhead{Common} & \nocolhead{Common} & \colhead{[$\mu$G]}  & \colhead{[$\mu$G]}  
}
\decimalcolnumbers
\startdata
PKS 1127$-$145&Core&&4.15E+03&3.92E$-$06&11.16&&5&8.9&9.50E+05&\\
&K-A&1.2&1.3&$0.3\mathrm{^b}$&11.16&11.2&&&6.2&0.42\\
&K-B&0.82&16&$0.3\mathrm{^b}$&11.16&18.6&&&13&0.25\\
&K-C&0.86&17&$0.3\mathrm{^b}$&11.16&28.5&&&13&0.17\\
3C 263&Core&&161&4.66E$-$06&1.98&&$30\mathrm{^b}$&9.4&8.86E+05&\\
&HS-K&&570&$0.3\mathrm{^b}$&$3\mathrm{^b}$&16.3&&&71&0.23\\
&L-NW&&190&8.0&$1\mathrm{^b}$&27.6&&&6.85&0.13\\
&L-SE&&44&8.0&$1\mathrm{^b}$&12.7&&&4.51&0.29\\
PKS 0637$-$752&Core&&6.30E+03&4.67E$-$06&17.91&&3.2&9.4&5.22E+05&\\
&K&0.8&48&$0.3\mathrm{^b}$&17.91&8.7&&&9.81&0.28\\
3C 273&Core&&3.20E+04&1.20E$-$05&10.48&&5&8.7&4.98E+05&\\
&K-A&0.85&133&$0.3\mathrm{^b}$&10.48&13&&&18&0.27\\
&K-C1&0.73&147&$0.3\mathrm{^b}$&10.48&16.8&&&18&0.55\\
&K-C2&0.75&301&$0.3\mathrm{^b}$&10.48&17.7&&&22&0.52\\
&K-B1&0.82&105&$0.3\mathrm{^b}$&10.48&14.2&&&17&0.65\\
&K-D1&0.77&424&$0.3\mathrm{^b}$&10.48&18.9&&&25&0.49\\
&K-DH&0.85&1.30E+03&$0.3\mathrm{^b}$&10.48&19.8&&&34&0.46\\
3C 380&Core&&7.45E+03&4.56E$-$06&12.75&&4.5&9.9&7.25E+05&\\
&K-K1&&163&$0.3\mathrm{^b}$&12.75&0.7&&&18&4.49\\
3C 454.3&Core&&1.22E+04&4.21E$-$06&7.28&&7.9&9.2&1.43E+06&\\
&HS-B&&216&59&$3\mathrm{^b}$&5.1&&&59&1.18\\
B2 0738+313&Core&&3.40E+03&4.71E$-$06&6.25&&9.2&9.4&9.12E+05&\\
&K-A&&4.2&$0.3\mathrm{^b}$&6.25&14.3&&&10&0.30\\
&HS-B1&&8.6&$0.3\mathrm{^b}$&$3\mathrm{^b}$&31.6&&&21&0.14\\
&HS-B2&&4.3&$0.3\mathrm{^b}$&$3\mathrm{^b}$&37.7&&&17&0.11\\
1928+738&Core&&3.60E+03&7.30E$-$06&5.24&&11&8.9&6.54E+05&\\
&K-A&&&$0.3\mathrm{^b}$&5.24&2.6&&&14&1.78\\
3C 371&Core&&1.20E+03&3.30E$-$05&11.45&&5&10.1&9.31E+04&\\
&K-A&0.76&37&$0.3\mathrm{^b}$&11.45&3.1&&&14&1.05\\
&K-B&0.73&15&$0.3\mathrm{^b}$&11.45&1.9&&&11&1.71\\
PKS 0521$-$365&Core&&7.61E+03&3.08E$-$05&1.67&&16&8.7&6.51E+05&\\
&K-A&0.6&150&$0.3\mathrm{^b}$&1.67&1.8&&&81&5.60\\
2201+044&Core&&1.68E+02&6.08E$-$05&6.39&&30&8.9&1.34E+05&\\
&K-$\alpha$&&&$0.3\mathrm{^b}$&1.86&1.1&&&28&5.43\\
&K-$\beta$&&&$0.3\mathrm{^b}$&1.86&1.6&&&27&3.74\\
&K-A&&&$0.3\mathrm{^b}$&1.86&2.2&&&37&2.72\\
\enddata
\tablecomments{$\alpha_{\rm R}$:\,radio flux index at 5GHz; $f_{\rm 5}$:\,radio flux density at 5GHz; $\theta$:\,radius of the emitting region; $\delta$:\,doppler beaming factor; $\theta_{\rm D}$:\,angular distance from the core; $\theta_{\rm inc}$:\,viewing angle; $\log (M_{\rm BH}/M_{\odot}$):\,black hole mass; $B_{\rm eq}$:\,equipartition magnetic field; $B_{\rm est}$:\,estimated magnetic field \\ $\mathrm{^a}$\ ``K" means knot, ``HS" means hotspot, and ``L" means lobe. \\$\mathrm{^b}$\ Assumed to be a listed value. }
\end{deluxetable*}

\begin{deluxetable*}{cccccccc}
\tablenum{3}
\tablecaption{List of the jet component's properties of Cen A \label{tab:Cen_A_data}}
\tablewidth{0pt}
\tablehead{
 \colhead{Component}  & \colhead{Observation Date} & \colhead{$f_{\rm 5}$} &  \colhead{$\theta$} & \colhead{$\delta$} & \colhead{$B_{\rm eq}$}  & \colhead{$d$} & \colhead{$\mathrm{Ref^b}$} \\
 \nocolhead{Common}  & \nocolhead{Common} & \colhead{[mJy]} & \colhead{[arcsec]} &  \nocolhead{Common} & \colhead{[$\mu$G]}  & \colhead{[pc]} & \nocolhead{Common}
}
\decimalcolnumbers
\startdata
\multicolumn{8}{c}{$\mathrm{Inner jet^a}$}\\
\hline
 J8 & 2008/11/27 & 502  & 5.9E-04   & 1.22 &9.0E+04 &0.31 & 1 \\
    & 2011/04/01 & 339  & 1.25E-03  & 1.22 &4.2E+04 &0.50 & 1 \\
 J7 & 2008/11/27 & 413  & 8.3E-04   & 1.22 &6.4E+04 &0.48 & 1 \\
    & 2010/07/24 & 502  & 1.66E-03  & 1.22 &3.7E+04 &0.61 & 1 \\
    & 2011/04/01 & 310  & 1.25E-03  & 1.22 &4.1E+04 &0.66 & 1 \\ 
 J5 & 2008/11/27 & 295  & 7.2E-04   & 1.22 &6.5E+04 &1.25 & 1 \\
 J4 & 2008/11/27 & 177  & 5.9E-04   & 1.22 &6.7E+04 &1.39 & 1 \\
    & 2009/09/05 & 221  & 1.15E-03  & 1.22 &4.0E+04 &1.46 & 1 \\
 J3 & 2008/06/09 & 59.0 & 1.6E-03   & 1.22 &2.1E+04 &1.77 & 1 \\
    & 2010/07/24 & 44.3 & 1.7E-03   & 1.22 &1.8E+04 &2.13 & 1 \\
 J2 & 2008/11/27 & 59.0 & 5.75E-04  & 1.22 &5.0E+04 &2.18 & 1 \\
 \hline
 \multicolumn{8}{c}{Lobe}\\
 \hline
 Region1&        & 2.9E+04   & 1.1E+03     & $\mathrm{1^c}$ & 1.36    &5.13E+05 & 2 \\
 Region2&        & 2.5E+04   & 9.7E+02     & $\mathrm{1^c}$ & 1.50    &3.24E+05 & 2 \\
 Region4&        & 1.0E+05  & 1.1E+03     & $\mathrm{1^c}$ & 2.00    &1.97E+05 & 2 \\
 Region5&        & 4.5E+04   & 1.0E+03     & $\mathrm{1^c}$ & 1.72    &3.76E+05 & 2 \\
\enddata
\tablecomments{$f_{\rm 5}$:\,radio flux density at 5GHz; $\theta$:\,radius of the emitting region; $\delta$:\,doppler beaming factor;  $B_{\rm eq}$:\,equipartition magnetic field; $d$:\,distance from the central black hole \\
$\mathrm{^a}$\ The knots of the inner jet are only listed with FWHM larger than beam size.\\
$\mathrm{^b}$\ 1.M{\"u}ller et al. (2014) 2.\citet{2009MNRAS.393.1041H}\\
$\mathrm{^c}$\ Assumed to be a listed value.
}
\end{deluxetable*}

\clearpage
\bibliography{sample63}{}
\bibliographystyle{aasjournal}

\end{document}